\documentclass{osa-article}

\journal{osajournal}
\usepackage{multirow}
\usepackage{bm}

\articletype{article}
\begin{document}
\title{Efficient Raman lasing and Raman-Kerr interaction in an integrated silicon carbide platform}

\author{Jingwei Li\authormark{1+}, Ruixuan Wang\authormark{1+}, Adnan A. Afridi\authormark{2}, Yaoqin Lu\authormark{2}, Xiaodong Shi\authormark{2}, Wenhan Sun\authormark{1}, Haiyan Ou\authormark{2}, and Qing Li\authormark{1}}

\address{\authormark{1}Department of Electrical and Computer Engineering, Carnegie Mellon University, Pittsburgh, PA 15213, USA}
\address{\authormark{2} DTU Electro, Technical University of Denmark, DK-2800 KGS. Lyngby, Denmark}
\address{\authormark{+} These authors contributed equally to this work.}
\email{\authormark{*}qingli2@andrew.cmu.edu} 


\begin{abstract}
Implementing stimulated Raman scattering in a low-loss microresonator could lead to Raman lasing. Here, we report the demonstration of an efficient Raman laser with $>50 \%$ power efficiency in an integrated silicon carbide platform for the first time. By fine tuning the free spectral range (FSR) of 43-$\mu$m-radius silicon carbide microresonators, the Stokes resonance corresponding to the dominant Raman shift of $777\ \text{cm}^{-1}$ ($23.3$ THz) is aligned to the center of the Raman gain spectrum, resulting in a low power threshold of $2.5$ mW. The peak Raman gain coefficient is estimated to be ($0.75 \pm 0.15) \ \text{cm}/\text{GW}$ in the 1550 nm band, with an approximate full width at half maximum of ($120 \pm 30$) GHz. In addition, the microresonator is designed to exhibit normal dispersion at the pump wavelength near 1550 nm while possessing anomalous dispersion at the first Stokes near 1760 nm. At high enough input powers, a Kerr microcomb is generated by the Stokes signal acting as the secondary pump, which then mixes with the pump laser through four-wave mixing to attain a wider spectral coverage. Furthermore, cascaded Raman lasing and occurrence of multiple Raman shifts, including $204\ \text{cm}^{-1}$ ($6.1$ THz) and $266\ \text{cm}^{-1}$ ($8.0$ THz) transitions, are also observed. Finally, we show that the Stokes Raman could also help broaden the spectrum in a Kerr microcomb which has anomalous dispersion at the pump wavelength. Our example of a 100-GHz-FSR microcomb has a wavelength span from 1200 nm to 1900 nm with 300 mW on-chip power.   
\end{abstract}
\noindent 
\section{Introduction}
The ubiquitous Raman effect, in which the incident photons experience inelastic scattering from the optical phonons of the matter, plays an important role in a range of applications such as material analysis \cite{Raman_book}, sensing \cite{Raman_review1}, optical communication \cite{Vahala_silica_Raman, Intel_Raman_laser_nature, Intenl_Raman_laser2, Noda_Raman_nature, Si_Raman_tunable}, and quantum information processing \cite{Raman_quantum_memory}. By implementing the stimulated Raman scattering in a low-loss microresonator, efficient Raman lasing can be achieved when the internal Raman gain exceeds the round-trip loss, thereby extending the wavelength range of conventional laser sources. This scheme has been demonstrated in silicon and silica microresonators with sub-mW power threshold and up to $45\%$ power efficiency \cite{Vahala_silica_Raman2, Noda_Raman_nature}. In the past decade, Raman lasing was also explored in wide-bandgap integrated photonic platforms such as diamond \cite{Loncar_diamond_Raman}, aluminum nitride \cite{AlN_Raman}, and lithium niobate \cite{Lonca_LN_Raman}. While these materials exhibit well-defined Raman peaks due to their crystalline structure, the reported power threshold is typically on the order of 10 mW or higher with an external power efficiency of less than $50\%$ (see Table 1). 

Recently, silicon carbide (SiC) emerged as a promising photonic and quantum material due to its unique properties, including strong Kerr nonlinearity (up to four times of silicon nitride) and hosting of various intrinsic and extrinsic color centers with appealing quantum properties \cite{Vuckovic_SiC_review, Li_SiC_Kerr_coeff, SiC_quantum_review}. These features, combined with the demonstration of the low-loss SiC-on-insulator (SiCOI) platform \cite{Noda_4HSiC_PhC, Vuckovic_4HSiC_nphoton, Ou_4HSiC_combQ}, have resulted in a range of competitive device applications, including Kerr microcombs \cite{Li_4HSiC_comb, Vuckovic_4HSiC_soliton, OuXin_soliton}, gigahertz-level electro-optic modulators \cite{Li_SiC_EOM}, as well as single and entangled photon sources \cite{Stefania_SiC_singlephoton, Awschalom_SiC_qubit, Li_SiC_entangled}. In 4H-SiC, the dominant Raman transition is around 777 $\text{cm}^{-1}$ (Fig.~1(b)) \cite{SiC_Raman_data}, which in terms of frequency shift ($\approx 23.3$ THz) is only second to that in diamond (see Table 1). To date, only Raman lasing corresponding to a frequency shift of $6.1$ THz (204 $\text{cm}^{-1}$) was reported in 4H-SiC, though the realized power efficiency was well below $1\%$ \cite{Ou_4HSiC_combQ}. 

In this work, we report the demonstration of an efficient Raman laser in an integrated 4H-SiCOI platform for the first time. To attain a high Raman efficiency, we work with over-coupled SiC microresonators with normal dispersion around the pump wavelength (Fig.~1(c)). To lower the power threshold, we fabricate SiC microrings with a nominal ring radius of 43 $\mu$m, which is varied by a step of 40 nm to align the Stokes resonance to the center of the Raman gain spectrum. In such compact microresonators, the fundamental transverse-electric mode has an intrinsic quality factor ($Q$) around $3.5$ million and a loaded $Q$ around $1.0$ million at 1550 nm (Fig.~1(d)). At an approximate input power of 10 mW, a strong Raman signal is observed at a frequency shift of $23.3$ THz away from the pump, displaying an estimated power efficiency of $51\%$ (Fig.~1(e)). When further increasing the pump power, a Kerr microcomb resulting from the anomalous dispersion near the Stokes is produced, which then mixes with the pump laser through four-wave mixing to reach an even broader spectrum. A detailed study of the Raman-Kerr interaction is carried out, revealing rich physics for such a compact nonlinear system. 
\begin{figure}[ht]
\centering
\includegraphics[width=0.55\linewidth]{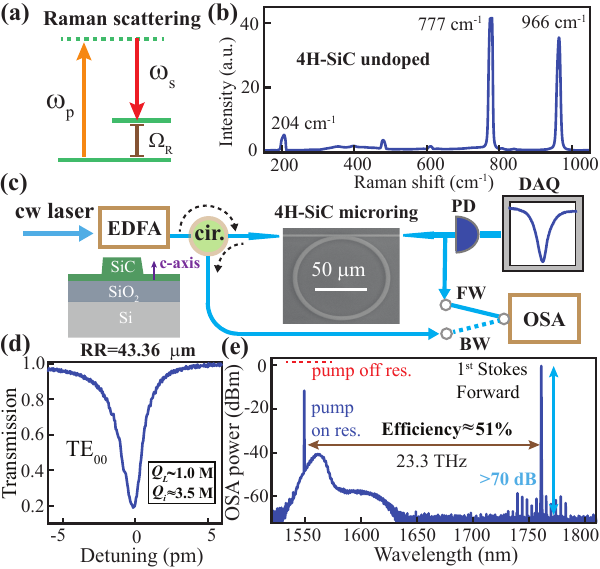}
\caption{(a) Energy diagram of the Stokes Raman process: $\omega_p$, $\omega_s$ and $\Omega_R$ represent the frequencies of the pump, Stokes, and Raman shift, respectively. (b) Raman spectroscopic data of an undoped, semi-insulating 4H-silicon carbide (SiC) wafer; (c) Experimental schematic: an amplified continuous-wave (CW) pump is coupled to the 4H-SiC chip through lensed fibers, whose forward (FW) and backward (BW) transmission and optical spectrum are recorded; (d) Representative linear transmission of the fundamental transverse-electric (TE$_{00}$) mode in a $43$-$\mu$m-radius microring at 1550 nm; and (e) Optical spectrum of the forward transmission for an approximate on-chip pump power of 10 mW (the red dashed line is the recorded power in OSA when tuning the pump laser off resonance). EDFA: erbium-doped fiber amplifier; PD: photodetector; DAQ: data acquisition; and OSA: optical spectrum analyzer.}
\label{Figure1}
\end{figure}

\begin{table}[ht]
\centering
\begin{tabular}{c c c c c c}
\hline
\multirow{2}{*}{\textbf{References}}& \multirow{2}{*}{\textbf{Materials}}& \textbf{Dominant Raman} & \textbf{Gain coeff.} & \textbf{Threshold} & \textbf{Power} \\
 & &\textbf{shift (THz)} & \textbf {($\text{cm}/\text{GW}$)} & \textbf{(mW)}&\textbf{efficiency ($\%$)} \\ 
 \hline
Latawiec et.al. (2015) \cite{Loncar_diamond_Raman} &Diamond& $\approx 40$ & 2.5 & 85 &  $<1$\\
\hline
Liu et.al.(2017) \cite{AlN_Raman}&AlN&$\approx 19$ & $0.25-0.45$& 8 & $10-15$\\
\hline
Yu et.al. (2020)\cite{Lonca_LN_Raman}&LN&$\approx 7.5$& 1.3 &20 & $\approx 42$\\
\hline
\textbf{This work}&\textbf{4H-SiC} & $\approx$ \textbf{23.3} & $\mathbf{0.75\pm 0.15}$ & \textbf{2.5} & \textbf{51}\\
\hline
 \end{tabular}
 \caption{Comparison of Raman lasing in crystalline, wide-bandgap integrated photonic platforms. Note that we have converted the ``slope efficiency'' defined in some references to the absolute power efficiency, which denotes the power ratio between the Raman signal and the pump.}
\label{Table1}
\end{table}

\section{Efficient Raman lasing and Raman gain coefficient measurement}
\begin{figure}[h]
\centering
\includegraphics[width=0.7\linewidth]{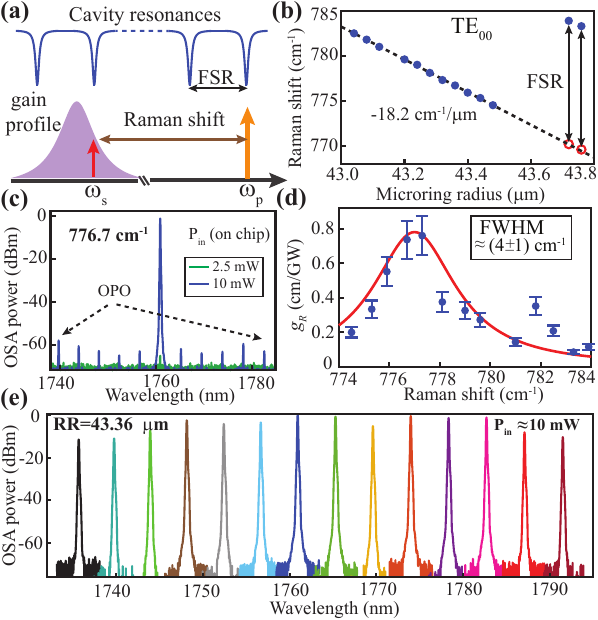}
\caption{(a) Illustration of the Stokes Raman process in a microresonator; (b) Experimentally measured Raman shift (solid circles) and a linear fit (dashed line) as a function of the microring outer radius (ring width fixed at $2.5\ \mu$m). (c) Superimposed optical spectra near the $777\ \text{cm}^{-1}$ Stokes signal at the Raman threshold power of 2.5 mW and maximum efficiency of 10 mW. The latter also results in optical parametric oscillation with the Stokes acting as the secondary pump (other spikes are the amplified spontaneous emission from EDFA transferred to the Stokes region through four-wave mixing); (d) Extracted Raman gain coefficients (markers with error bars) corresponding to different Raman shifts. The red solid line is a Gaussian fit with its full width at half maximum (FWHM) estimated to be ($120\pm 30$) GHz. (e) Superimposed Stokes corresponding to the Raman shift of $776.7\ \text{cm}^{-1}$ by varying the pump resonance in the 1529-1572 nm range for a fixed on-chip power of 10 mW.}
\label{Figure2}
\end{figure}
As illustrated in Fig.~2(a), the effective Raman gain and frequency shift are both determined by the relative position of the Stokes resonance within the Raman gain profile. Hence, our design efforts are focused on aligning the Stokes resonance to the peak Raman gain for a low power threshold and achieving over coupling for a high Raman efficiency. In the literature, most of the existing experiments relied upon either random frequency matching \cite{AlN_Raman, Loncar_diamond_Raman} or using a microresonator whose FSR is smaller than the Raman gain bandwidth \cite{Lonca_LN_Raman}, both of which would result in an inflated power threshold. Here, we adopt a different approach by employing a compact microresonator (nominal radius near $43\ \mu$m with a corresponding FSR around 400 GHz) and varying its radius from $43\ \mu$m to $43.76\ \mu$m in 40 nm increments (20 microrings in total). A straightforward calculation predicts an incremental frequency shift of $-21.7$ GHz  ($\approx \frac{-0.04}{43} \times 23.3$ THz) for the $777\ \text{cm}^{-1}$ Stokes resonance in each 40 nm increase of the ring radius. A total of 20 microrings therefore is able to cover the whole FSR, ensuring that there is at least one microring within $21.7$ GHz of the peak Raman gain. 

To obtain strong coupling at the pump and Stokes resonances, we resort to the straight coupling scheme, where a 900-nm-wide straight waveguide is evanescently coupled to the 43-$\mu$m-radius SiC microring with a 200 nm gap. The linear transmission measurement as shown in Fig.~1(d) confirms that a coupling $Q$ of $1.4$ million (versus an intrinsic $Q$ of $3.5$ million) is attained near the pump wavelength of 1550 nm. The coupling $Q$ of the $777\ \text{cm}^{-1}$ Stokes resonance is expected to be smaller, which is typical for straight waveguide coupling due to increased modal overlap at longer wavelengths. In fact, numerical simulation points to a coupling $Q$ of $0.6$ million near 1760 nm (see Supplementary). This would lead to a loaded $Q$ around $0.5$ million of the Stokes resonance if the intrinsic $Q$ is the same as 1550 nm. 

With the optimized design, we proceed to fabricate the devices using the standard nanofabrication process which is described in detail in Ref.~\citeonline{Li_4HSiC_comb}. Briefly, we begin with a 4H-SiCOI chip with 700 nm SiC on top of 2 $\mu$m oxide (NGK Insulators, LTD.). The pattern is first defined using e-beam lithography and subsequently transferred to the SiC layer using plasma etching for an etch depth around 575 nm, leaving an approximate 125 nm of pedestal layer (i.e., unetched SiC). After cleaning, $2\ \mu$m oxide clad  is deposited on the 4H-SiC layer to encapsulate the devices. Next, we characterize the Raman chip using the experimental schematic described in Fig.~1(c), where both forward and backward transmissions are measured. The fiber-to-chip coupling is achieved by implementing inverse tapers on the SiC chip, whose coupling loss is  estimated to be $3$-$4$ dB per facet (see Supplementary for more information). The overall fiber-to-fiber insertion loss of this chip, however, is typically around $10$-$14$ dB, as this chip suffers from relatively strong charging effects in e-beam lithography that left small unexposed areas (cracks) in random places of waveguides and microresonators (see Supplementary for details). When appearing in waveguides, each crack introduces an approximate 2 dB additional loss. The presence of the charging-induced crack in a microring, on the other hand, will render the device useless since there is typically no resonance due to extremely low intrinsic $Q$s. 

Figure 2(b) summarizes the measured Raman shifts for 13 microrings with different radii (the missing data points are caused by the charging effects appearing in the corresponding microrings except for the ring radius of $43.6\ \mu$m, which is discussed in Fig.~4(b)), most of which follow the predicted linear frequency shift with the increased ring radius. The Raman shifts of the last two data points, corresponding to radii of $43.72\ \mu$m and $43.76\ \mu$m, are one FSR higher than the predicted trend. This can be understood as when the Stokes resonance is gradually moved toward the pump (by reducing the FSR), its adjacent resonance with a lower azimuthal order becomes closer to the center of the Raman gain and lases instead. 

Among all the working microresonators, the one with the $43.36\ \mu$m radius exhibits the lowest Raman threshold of $2.5$ mW with a measured Raman shift of $776.7\ \text{cm}^{-1}$ (Fig.~2(c)). For each Raman shift, we estimate the corresponding Raman gain coefficient $g_R$ based on the knowledge of the power threshold $P_\text{th, Raman}$ \cite{Vahala_silica_Raman}: 
\begin{equation}
P_{\text{th, Raman}}\approx \frac{\pi^2n_\text{SiC}^2V_{\text{eff}}}{\lambda_p\lambda_s g_R}\cdot\frac{Q_{c,p}}{Q_{l,p}^2}\cdot\frac{1}{Q_{l,s}},
\label{Eq_Pth}
\end{equation}
where $n_\text{SiC}$ is the refractive index of SiC ($n_\text{SiC}\approx 2.6$); $V_\text{eff}$ is the effective mode volume ($V_\text{eff}\approx 270\ {\mu m}^3$ for 43-$\mu$m-radius SiC microrings); $\lambda_p$ and $\lambda_s$ denote the wavelengths of the pump and Stokes, respectively; $Q_{c,p}$ and $Q_{l,p}$ are the coupling $Q$ and loaded $Q$ of the pump resonance, respectively; and $Q_{l,s}$ is the loaded $Q$ of the Stokes resonance. In Eq.~\ref{Eq_Pth}, the only parameter that cannot be directly measured is $Q_{l,s}$, which is due to a lack of a tunable laser source near the Stokes resonance. However, as explained earlier, we can infer the coupling $Q$ at the Stokes resonance based on the measured coupling $Q$ at 1550 nm, while assuming that the intrinsic $Q$ at the Stokes resonance is the same as the pump. Take the pump resonance shown in Fig.~1(d) (which corresponds to the microring with the lowest power threshold) for example, its coupling $Q$ and loaded $Q$ at 1760 nm are approximated to be $0.6$ million and $0.5$ million, respectively. These numbers can be corroborated using the fact that when the Stokes power is large enough, optical parametric oscillation (OPO) is observed near the Stokes resonance because of its anomalous dispersion (Fig.~2(c)). (Note that the OPO pairs are generated 5 FSRs away from the Stokes while the rest peaks observed in Fig.~2(c) are the transferred noise from the EDFA.) The OPO threshold with the Stokes serving as the secondary pump is given by \cite{Vuckovic_4HSiC_soliton}: 
\begin{equation}
P_{\text{OPO}}\approx \frac{\pi n_g^2 V_{\text{eff}}}{4\lambda_s n_2}\cdot\frac{Q_{c,s}}{Q_{l,s}^3},
\label{Eq_Popo}
\end{equation}
where $n_g$ ($n_g\approx 2.7$) is the group index and $n_2$ is the Kerr nonlinear index of 4H-SiC ($n_2\approx 9.1\times 10^{-19} \text{m}^2/\text{W}$ for the TE polarization) \cite{Li_SiC_Kerr_coeff}. The computed power threshold of $5.0$ mW is consistent with the experimental data shown in Fig.~2(c) (the on-chip Stokes power corresponding to 10 mW input is approximately $10\times 51\%=5.1$ mW). Hence, recording the OPO threshold power for each Raman shift provides an effective calibration for the estimation of the Raman gain coefficient. In Fig.~2(d), we plot the extracted Raman gain coefficient as a function of the corresponding Raman shift, where a Gaussian fit reveals a full width at half maximum (FWHM) around $(4\pm 1)\ \text{cm}^{-1}$, or $(120\pm30)$ GHz. 

In addition to the efficient Raman lasing, the Stokes is expected to shift in sync with the pump resonance to maintain a fixed Raman shift. Figure 2(e) shows one such example for the microring with radius of $43.36\ \mu$m, where we sequentially select pump resonances from 1529 nm to 1572 nm (limited by the EDFA range) for a fixed on-chip power of 10 mW and superimpose the resultant Stokes generated from 1735 nm to 1792 nm. The non-uniformity in the Stokes power is mainly attributed to the slight variations in the input power and resonant properties of different azimuthal orders. Nevertheless, the result displayed in Fig.~2(e) demonstrates the ease of wavelength tuning for the Raman lasing process, which is important for many practical applications. 

\section{Raman-Kerr interaction and comb generation}
\begin{figure}[ht]
\centering
\includegraphics[width=0.8\linewidth]{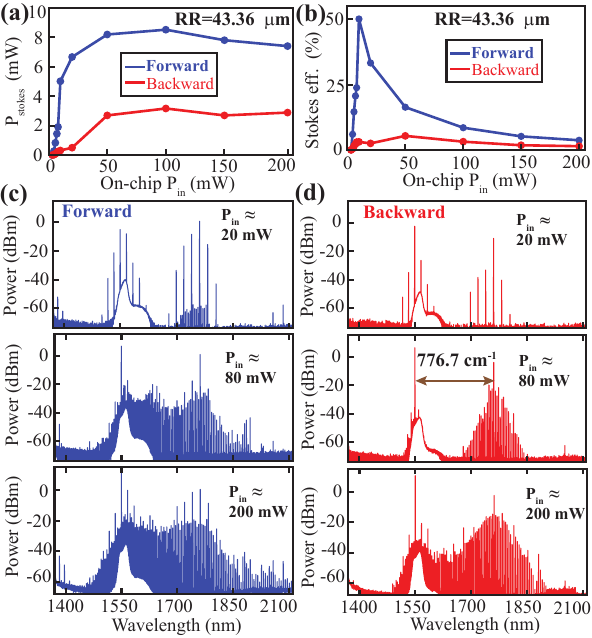}
\caption{(a) Inferred on-chip Stokes power of the $776.7\ \text{cm}^{-1}$ Raman shift for the forward and backward transmission after accounting for coupling losses; (b) Stokes efficiency by normalizing the on-chip Stokes power by the pump power; (c) and (d) Measured optical spectra for the forward and backward transmission at three different pump powers.}
\label{Figure3}
\end{figure}
As we further increase the pump power, the Stokes signal saturates to a level of 6-8 mW on chip for the forward transmission (Fig.~3(a)). Hence, the maximum Stokes efficiency is achieved when the input power is around 10 mW, beyond which it begins to decrease with increased powers (Fig.~3(b)). This is not surprising since when the Stokes power is above the OPO threshold (around 5 mW), it can act as a secondary pump for the Kerr microcomb generation due to the anomalous dispersion around the Stokes resonance (see Supplementary). In addition, the Raman process in general would result in clockwise (backward) and counter-clockwise (forward) Stokes inside the microresonator \cite{Lonca_LN_Raman}, which might have slightly different power thresholds. To understand the interactive dynamics between Raman and the pump laser, we plot the optical spectra corresponding to the forward and backward transmissions at three representative powers in Fig.~3(c). As can be seen, a primary comb with multiple-FSR-separation is first generated near the Stokes resonance at a pump power of 20 mW (which can be compared to Fig.~2(c) to support the OPO identification). These comb lines are also transferred to the pump region due to a non-degenerate four-wave mixing process (so-called four-wave mixing Bragg scattering): $\omega_{\textmd{comb near pump}}=\omega_\textmd{comb near Stokes}+\omega_p-\omega_s$ \cite{Li_FWMBS}. In fact, the small peaks near the Stokes resonance are the transferred noise from the EDFA (amplified spontaneous emission) based on the same principle: $\omega_{\textmd{noise near Stokes}}=\omega_\textmd{noise near pump}+\omega_s-\omega_p$. The backward spectrum at 20 mW consists of the reflection of the pump and the backward-propagating Stokes comb generated inside the microring. As we increase the pump power to 80 mW, we begin to see major differences in the forward and backward optical spectra: the backward spectrum contains a filled Kerr microcomb (termed as Raman-Kerr comb for simplicity) that is well separated from the 1550 nm pump, while the forward spectrum consists of additional combs lines between the pump and the Stokes resonance. Such difference can be understood as the forward optical spectrum consists of the counter-clockwise Raman-Kerr comb and the pump laser, both of which propagate in the same direction inside the microresonator and hence interact with each other through four-wave mixing. In contrast, the backward spectrum consists of the reflected portion of the pump and the clockwise Raman-Kerr comb without them mixing with each other inside the microresonator. Finally, as the pump power is increased to 200 mW, both the forward and backward spectra are significantly widened, though the forward spectral coverage is modestly wider, spanning from 1400 nm to 2100 nm.  
\begin{figure}[ht]
\centering
\includegraphics[width=0.8\linewidth]{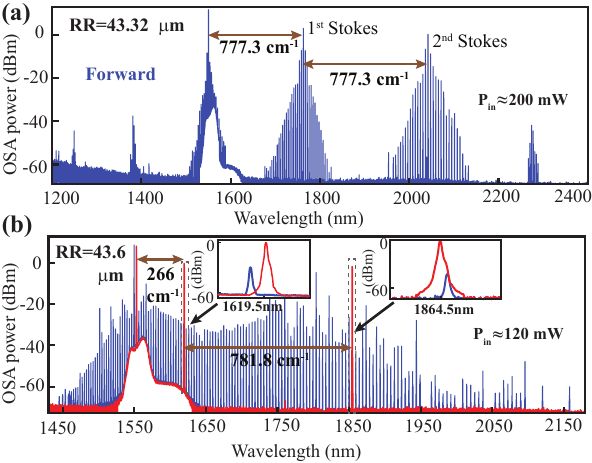}
\caption{(a) Cascaded Raman shifts of $777.3 \text{cm}^{-1}$ from the microring with outer radius of $43.32\ \mu$m at an input power of 200 mW. (b) The blue solid line is the Raman-Kerr comb from the microring with outer radius of $43.6\ \mu$m when pumped at the TE$_{00}$ resonance near $1549.5$ nm. The red solid line corresponds to pumping at the next resonance near $1552.9$ nm, whose spectrum consists of two cascaded Raman shifts, one near $266\ \text{cm}^{-1}$ and the other near $781.8\ \text{cm}^{-1}$, without Kerr comb. The insets show the zoom-in spectra near the two Raman shifts, suggesting that the Stokes are likely formed by a different mode family (TM).}
\label{Figure4}
\end{figure}

In addition to Stokes-induced Kerr microcomb, we also observe cascaded Raman generation in a few microrings, pointing to even more complicated interactions between the Raman and Kerr effects \cite{Intenl_Raman_laser2, Ou_4HSiC_combQ}. For example, Fig.~4(a) shows that there are two cascaded $777.3\ \text{cm}^{-1}$ Stokes in the microring with a radius of $43.32\ \mu$m, each generating a Kerr microcomb in the adjacent region. Moreover, while the $777\ \text{cm}^{-1}$ Stokes dominates over other Raman transitions in most of the microring resonators, we also observe different Raman shifts in certain geometries. For example, in the microring with a radius of $43.6\ \mu$m radius, when pumping the TE$_{00}$ resonance near $1549.5$ nm, a regular Raman-Kerr comb is observed in the forward transmission (blue solid line in Fig.~4(b)). However, when we tune the pump laser to the $1552.9$ nm resonance, only two cascaded Raman shifts are observed in the spectrum, one corresponding to $266\ \text{cm}^{-1}$ and the other corresponding to $781.8\ \text{cm}^{-1}$ (red solid line in Fig.~4(b)). An inspection of the optical spectrum near the two Stokes (see the two insets in Fig.~4(b)) suggests that the two Stokes are not from the TE$_{00}$ mode family. Instead, they are likely the TM$_{00}$ modes of the microring that happen to be accidentally frequency matched to the TE$_{00}$ pump for the observed Raman transitions. This may explain why the $266\ \text{cm}^{-1}$ Raman transition is often absent in the bulk-wafer measurement (see Fig.~1(b)) \cite{SiC_Raman_data}.
 
\begin{figure}[ht]
\centering
\includegraphics[width=0.8\linewidth]{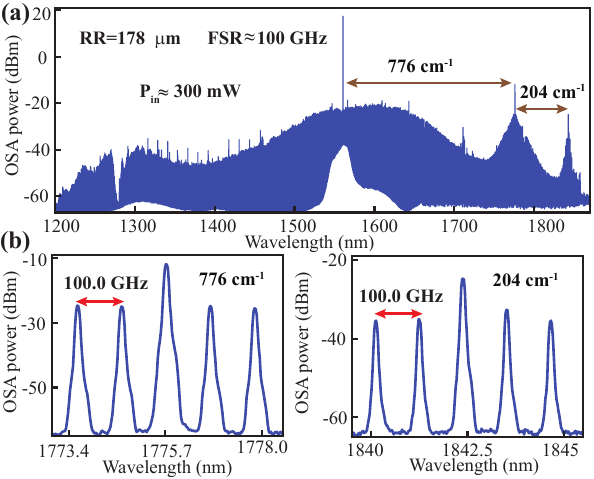}
\caption{(a) Optical spectrum of a SiC microring with a radius of 178 $\mu$m (corresponding free spectral range around 100 GHz), which consists of the Kerr comb from the pump itself and two cascaded Raman peaks near $776\ \text{cm}^{-1}$ and $204\ \text{cm}^{-1}$. (b) Zoomed-in spectra near the two Raman peaks.}
\label{Figure5}
\end{figure}

So far, we have only discussed the stimulated Raman scattering in SiC microresonators that exhibit normal dispersion at the pump wavelength, a preferred configuration to avoid competition between the Raman and Kerr effects at the same wavelength \cite{Gaeta_Raman_competition}. It is possible, however, to observe Stokes appearing in a Kerr microcomb as well. Figure 5 shows one such example for a 178-$\mu$m-radius SiC microring resonator. The TE$_{00}$ mode at this increased radius possesses a weak anomalous dispersion (see Supplementary for more details). When pumped at 300 mW power, a broad Kerr microcomb spanning from 1200 nm to 1900 nm is generated along with two cascaded Raman transitions, with one peak corresponding to $776 \text{cm}^{-1}$ and the other to $204\ \text{cm}^{-1}$. A zoom-in view of the two Raman peaks reveals that they are from the same mode family. 

Finally, we want to comment on the coherence properties of the Raman-Kerr comb generated in this work. Most of them, with the only exception of the primary comb in Fig.~3(c), are believed to be in the chaotic state that is not phase-locked to the pump laser. It is possible, however, to observe a soliton microcomb with the Stokes serving as the secondary pump or the coexistence of the Stokes soliton with the Kerr soliton generated by the pump \cite{Raman_Kerr_theory, Vahala_Stokes_soliton}. A detailed discussion on these aspects of Raman-Kerr interactions is beyond the scope of this paper and will be left to future works. 

\section{Conclusion}
In conclusion, we performed a thorough investigation of the Raman effect in low-loss 4H-SiC microresonators, resulting in the demonstration of an efficient Raman laser with $>50\ \%$ power efficiency and a detailed characterization of the Raman gain coefficient for the dominant $777\ \text{cm}^{-1}$ ($23.3$ THz) Stokes transition, both of which are the first in an integrated SiC platform (to the best of our knowledge). In addition, our study of the Stokes-induced Kerr comb and its interplay with the pump laser revealed the different roles of four-wave mixing in the forward and backward transmissions as well as the rich interactions between the Raman and Kerr effects. Finally, we also observed the occurrence of other Raman transitions such as $204\ \text{cm}^{-1}$ ($6.1$ THz) and $266\ \text{cm}^{-1}$ ($8.0$ THz) along with the $777\ \text{cm}^{-1}$ Stokes, a feature that can be utilized to broaden the Stokes-induced or pump-induced Kerr microcombs. We believe that our results fill up the gap in the understanding of the stimulated Raman process in 4H-SiC microresonators, which could potentially lead to a myriad of applications including efficient Raman lasers and novel approaches to generating broadband microcombs.  

\newpage
\title{Supplementary Material}
\setcounter{equation}{0}
\setcounter{figure}{0}
\setcounter{table}{0}
\setcounter{section}{0}
\renewcommand{\thetable}{S\arabic{table}}
\renewcommand{\thefigure}{S\arabic{figure}}
\renewcommand{\theequation}{S\arabic{equation}}
\noindent 
\section{Inverse tapers and charging effect}
\begin{figure}[ht]
\centering
\includegraphics[width=0.75\linewidth]{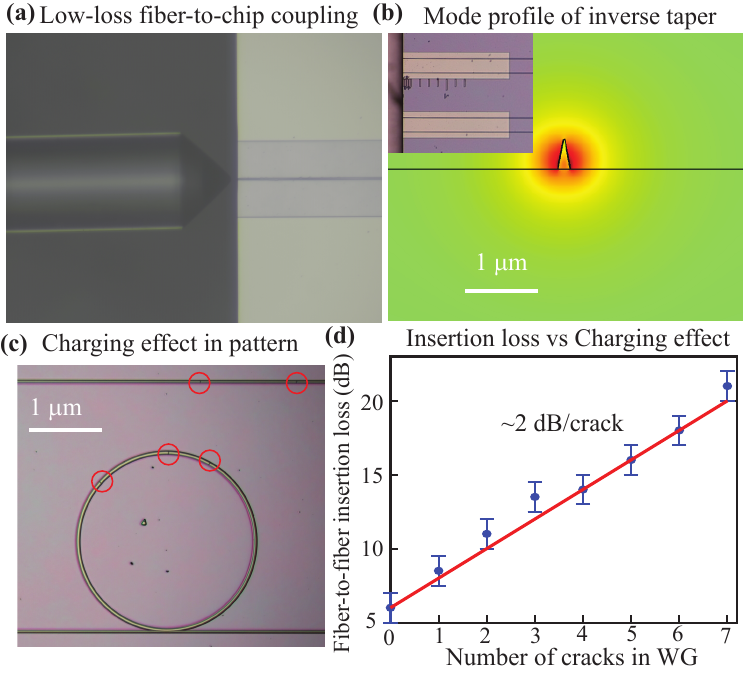}
\caption{(a) Optical micrograph of a lensed fiber aligned to a polished SiC facet; (b) The mode profile of the SiC inverse taper with a (bottom) width around 250 nm. The inset shows the optical micrograph of the taper region where the pedestal layer is removed by an additional step of lithography; (c) Optical micrograph highlighting charging-induced cracks (red circles) in waveguides and microresonators; and (d) Statistically averaged fiber-to-fiber insertion loss as a function of the number of cracks appearing in waveguides.}
\label{FigureS_taper}
\end{figure}

\noindent In this work, efficient fiber-to-chip coupling is achieved by aligning a lensed fiber (with a mode field diameter of $2.5\ \mu$m) to the SiC inverse taper implemented at the chip facet (Fig.~S1(a)). The mode profile of the inverse taper is provided in Fig.~S1(b), which has an estimated coupling loss of 2 dB by computing its modal overlap with the lensed fiber. Given that the presence of a pedestal layer hinders the mode expansion and thus reduces the coupling efficiency, we remove the pedestal layer in the taper region using an additional step of lithography (see the inset of Fig.~S1(b)). That is to say, in the microring region the SiC is 700 nm thick with 125 nm pedestal, while the SiC is only around 550 nm thick without pedestal in the taper region. We then polish the SiC facets so that the tip of the inverse tapers is only a few microns away from the facet. In our prior experiments, the inverse tapers would typically fail at high optical input powers due to residues from the polishing step that caused strong optical absorption \cite{Li_4HSiC_comb}. After switching to a more thorough cleaning process, the inverse tapers can now sustain optical powers up to several watts without getting burnt. 

When writing long waveguides in the e-beam lithography, the charging effect caused by charge accumulation in insulating substrates becomes more apparent, resulting in cracks in the writing area as highlighted in Fig.~1S(c). Though a few mitigating methods including evaporating a thin layer of aluminum on top of the e-beam resist have been tried, it is still common to get 1-3 cracks in a 5-mm-long waveguide. For dense patterns consisting of many devices (including this Raman chip), the charging effect can be even stronger. Fitting the measured insertion loss as a function of the observed cracks in various waveguides reveals a positive correlation between the increased losses and the number of charging-induced cracks. Statistically, each crack introduces an approximate $1.5$-$2$ dB additional loss. For waveguides that are free of the charging effect, the total chip insertion loss is only around 6-7 dB, suggesting a coupling loss around 3 dB for each inverse taper (which is slightly higher than the 2 dB number obtained from simulation). 

\section{Dispersion of 43-$\mu$m-radius SiC microrings}
\begin{figure}[ht]
\centering
\includegraphics[width=0.75\linewidth]{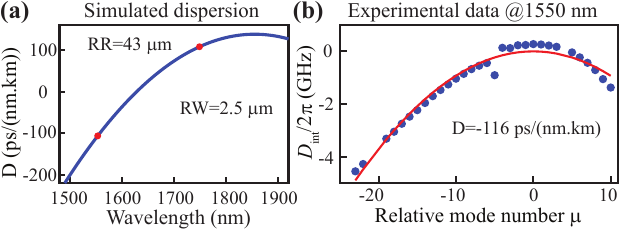}
\caption{(a) Simulated dispersion of the 43-$\mu$m-radius SiC microring used in this work: a ring width of $2.5\ \mu$m exhibits normal dispersion around 1550 nm and anomalous dispersion around 1760 nm (as highlighted by the two red dots in the figure); and (b) Experimentally measured dispersion by computing the integrated dispersion $D_\text{int}$ as a function of different azimuthal orders, confirming normal dispersion in the 1550 nm band. $D_\text{int}$ is defined as $D_\text{int} \equiv \omega_{\mu} - \omega_0 - D_1\mu$ where $\mu$ is the relative azimuthal order to the pump resonance (i.e., $\mu=0$ for the pump mode), $\omega_{\mu}$ is the corresponding resonance frequency, and $D_1$ is the free spectral range of the resonator.}
\label{FigureS_Dispersion_R43}
\end{figure}

\noindent As discussed in the main text, the 43-$\mu$m-radius SiC microring employed in this work is designed to exhibit normal dispersion near the pump wavelength (1550 nm) while possessing anomalous dispersion in the Stokes wavelength corresponding to the dominant $777\ \text{cm}^{-1}$ Raman shift (1760 nm). Numerical simulation performed in Fig.~S2(a) suggests that this can be achieved by choosing a ring width of $2.5\ \mu$m with a height of 700 nm and a pedestal layer of 125 nm. After fabricating these microrings, a linear dispersion measurement was performed to confirm the normal dispersion in the 1550 nm region (see one example in Fig.~S2(b)). 

\section{Waveguide-resonator coupling simulation}
\begin{figure}[ht]
\centering
\includegraphics[width=0.75\linewidth]{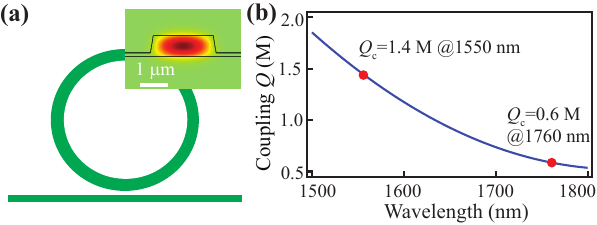}
\caption{(a) Schematic of straight waveguide coupling between a microring resonator and the access waveguide. The inset shows the mode profile of the 43-$\mu$m-radius SiC microring used in this work, which has a width of $2.5\ \mu$m and a height of $700$ nm (the pedestal layer is 125 nm). (b) Simulated coupling $Q$ as a function of wavelengths: the access waveguide has a width of 900 nm and a coupling gap of 200 nm.}
\label{FigureS_coupling}
\end{figure}

\noindent The straight coupling scheme, as illustrated in Fig.~S3(a), is ideal for achieving over-coupling in both the pump and Stokes resonances. This is because in this coupling scheme, the modal overlap factor plays a dominant role (given the short interaction length), which typically results in stronger coupling at longer wavelengths \cite{Li_FWMBS}. This means if we can achieve over-coupling in the 1550 nm band, the degree of over-coupling at 1760 nm should be even stronger (see Fig.~S2(b) for simulation). Typically, due to the limited accuracy in the dimensional control from nanofabrication, we have to adjust the gap by up to 50 nm to match the observed coupling $Q$ with the experimental data. In this case, there is good agreement between the measured coupling $Q$ at 1550 nm and the simulation data based on the designed gap (200 nm). Nevertheless, once we match the coupling $Q$ at the pump wavelength ($1.4$ million at 1550 nm), numerical simulation predicts an approximate coupling $Q$ of 0.6 million at 1760 nm. This number is also verified by comparing the predicted OPO power threshold with the experimental data (see main text). 

\section{Raman power efficiency estimation}
\begin{figure}[ht]
\centering
\includegraphics[width=0.6\linewidth]{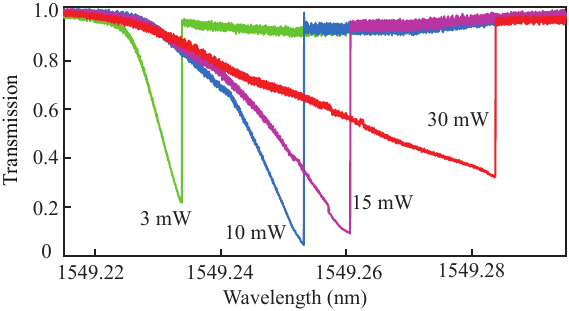}
\caption{Normalized pump transmission of the $43.36$-$\mu$m-radius SiC microring at different input powers.}
\label{FigureS_pump_tran}
\end{figure}

\noindent The Raman power efficiency around $51\ \%$ is estimated based on the OSA spectrum shown in Fig.~1(e), which is given by the ratio between the recorded Stokes power (when the pump is tuned into resonance) and the off-resonance pump power. This measurement assumes that the Stokes and pump experience the same out-coupling loss from the SiC chip (the numerical value of the insertion loss is less important as long as it stays the same between the two measurements). It is possible, however, that the coupling loss of the 1760 nm Stokes is slightly higher than that of the 1550 nm pump, as suggested by preliminary characterization carried out by the DTU team using a supercontinuum source on a different SiC chip. This would indicate a slightly higher power efficiency (by $0.5$ dB) if verified. To be conservative, the reported number here does not adjust for the difference of the insertion loss between the Stokes and pump. 

One may notice that the extinction ratio of the pump resonance shown in Fig.~1(e) ($\approx 14$ dB) is higher than what is suggested by the linear transmission in Fig.~1(d) ($\approx 7$ dB). This is confirmed by comparing the pump transmissions at various input powers in Fig.~\ref{FigureS_pump_tran}. The result is also consistent with the interpretation of the over-coupled pump resonance in the linear regime and high Raman conversion efficiency that depletes the pump when the input power is around 10 mW. 

\section{Additional information on 178-$\mu$m-radius SiC microrings}
\begin{figure}[ht]
\centering
\includegraphics[width=0.75\linewidth]{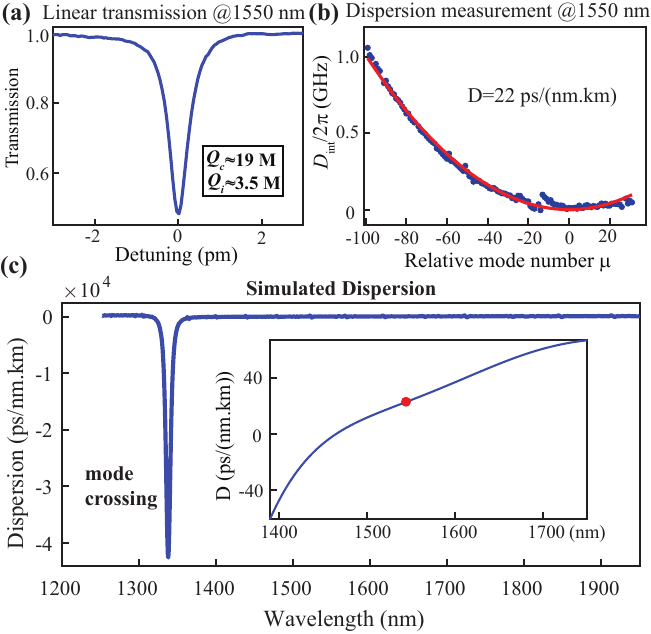}
\caption{(a) Representative linear transmission of the fundamental TE$_{00}$ mode of the 178-$\mu$m-radius SiC microring studied in Fig.~5 of the main text, showing an intrinsic $Q$ around $3.5$ million and a coupling $Q$ around 19 million. The microring has a ring width of $3.5\ \mu$m and an access waveguide with a width of 1450 nm and a pulley coupling length of 30 $\mu$m; (b) Dispersion measurement confirms anomalous dispersion in the 1550 nm band, which is also consistent with simulation shown in (c) at 1550 nm; and (c) Simulated dispersion for the TE$_{00}$ mode of the 178-$\mu$m-radius SiC microring, which shows a mode crossing around 1330 nm.}
\label{FigureS_R178}
\end{figure}

\noindent The 178-$\mu$m-radius SiC microring is from a different chip which was fabricated using a similar process but with different designs. Notably, it has the same SiC thickness of 700 nm and a pedestal layer of 125 nm. The most significant difference is that the coupling to the fundamental TE$_{00}$ mode is based on the pulley coupling instead of straight coupling \cite{Li_FWMBS}. The linear transmission shown in Fig.~S5(a) reveals that the TE$_{00}$ mode is under-coupled at 1550 nm, which explains why the Raman signals are much weaker compared to those observed in 43-$\mu$m-radius SiC microrings. The dispersion characterization in Fig.~S5(b) confirms that the larger-radius-microring exhibits weak but anomalous dispersion near 1550 nm, which is consistent with the simulated dispersion (see Fig.~S5(c)). For the 700 nm SiC with 125 nm pedestal, numerical simulation also points to a possible mode crossing between the TE$_{00}$ and TM$_{00}$ mode families around the wavelength of 1300 nm. This may explain the disappearance of certain comb lines in that region as shown in Fig.~5(a) in the main text.

\begin{backmatter}
\bmsection{Funding}
This work was supported by NSF (2127499, 2131162). The DTU team would like to acknowledge the support of the European Union's Horizon 2020 FET Open project (SiComb, No.~899679) and the Villum Fonden (Grant No.~50293).  

\bmsection{Acknowledgments}
The CMU team acknowledges use of the Materials Characterization Facility at Carnegie Mellon University supported by grant MCF-677785, as well as the equipment support from Dr.~Lijun Ma and Dr.~Oliver Slattery at NIST. R.~Wang also acknowledges the support of Tan Endowed Graduate Fellowship from CMU.

\bmsection{Disclosures}  The authors declare no conflicts of interest.

\bmsection{Data Availability} Data underlying the results presented in this paper are not publicly available at this time but may be obtained from the authors upon reasonable request.

\end{backmatter}


\bibliography{SiC_Ref}

\end{document}